\documentclass[pra,twocolumn,showpacs,superscriptaddress,floatfix,amsmath,nofootinbib,amssymb]{revtex4} 

\usepackage[english]{babel}
\usepackage{graphicx}
\usepackage{dcolumn}
\usepackage{bm}
\usepackage{verbatim}

\newcommand{\ket}[1]{\left| {#1} \right\rangle}
\newcommand{\bra}[1]{\left\langle {#1} \right|}
	
\newcommand{\braket}[2]{\left\langle {#1}\left|{#2}\right.\right\rangle}
\newcommand{\proj}[2]{\left| {#1} \right\rangle\!\left\langle {#2} \right|}

\newcommand{\tr}{\operatorname{Tr}}

\def\slashchar#1{\setbox0=\hbox{$#1$} 
\dimen0=\wd0 
\setbox1=\hbox{/} \dimen1=\wd1 
\ifdim\dimen0>\dimen1 
\rlap{\hbox to \dimen0{\hfil/\hfil}} 
#1 
\else 
\rlap{\hbox to \dimen1{\hfil$#1$\hfil}} 
/ 
\fi}

\begin{document}


\title{Population bound effects on bosonic correlations in non-inertial frames}
\author{Eduardo Mart\'{i}n-Mart\'{i}nez}%
 \email{martin@iff.csic.es}
\author{Juan Le\'on}
\email{leon@iff.csic.es}
 \homepage{http://www.imaff.csic.es/pcc/QUINFOG/}
\affiliation{%
Instituto de F\'{i}sica Fundamental, CSIC\\
Serrano 113-B, 28006 Madrid, Spain.\\
}


\date{April 13, 2010}

\begin{abstract}
We analyse the effect of bounding the occupation number of bosonic field modes on the correlations among all the different spatial-temporal regions in a setting in which we have a space-time with a horizon along with an inertial observer. We show that the entanglement between $A$ (inertial observer) and $R$ (uniformly accelerated observer) depends on the bound $N$, contrary to the fermionic case. Whether or not decoherence increases with $N$ depends on the value of the acceleration $a$. Concerning the bipartition $A\bar R$ (Alice with an observer in Rindler's region $IV$), we show that no entanglement is created whatever the value of $N$ and $a$. Furthermore, $AR$ entanglement is very quickly lost for finite $N$ and for $N\rightarrow\infty$. We will study in detail the mutual information conservation law found for bosons and fermions.Ê By means of the boundary effects associated to $N$ finiteness, we will show that for bosons this law stems from classical correlations while for fermions it has a quantum origin. Finally, we will present the strong $N$ dependence of the entanglement in $R\bar R$ bipartition and compare the fermionic cases with their finite $N$ bosonic analogs. We will also show the anti-intuitive dependence of this entanglement on statistics since more entanglement is created for bosons than for their fermion counterparts. 
\end{abstract}

\maketitle


\section{Introduction}

To analyse quantum correlations in non-inertial settings it is necessary to combine knowledge from different branches of Physics; quantum field theory in curved space-times and quantum information theory. This combination of disciplines became known as relativistic quantum information, which is developing at an accelerated pace \cite{Alsingtelep,TeraUeda2,ShiYu,Alicefalls,AlsingSchul,SchExpandingspace,Adeschul,KBr,LingHeZ,ManSchullBlack,PanBlackHoles,AlsingMcmhMil,DH,Steeg,Edu2, schacross,Hu}. It provides novel tools for the analysis of the Unruh effect \cite{DaviesUnr,Unruh,Takagi,Crispino} allowing us to study the behaviour of the correlations shared between non-inertial observers.

Let us consider a bipartite system (Alice-Rob), in which one of the partners (Rob) is undergoing a uniform acceleration and therefore describing the world which Rindler coordinates. As pointed out in \cite{AlsingSchul,Edu4} there are 3 possible bipartitions that can be considered when analysing entanglement in the previous setting; 1) The entanglement of the inertial observer with field modes in Rindler's region $I$ (Alice-Rob $AR$), 2) The entanglement of the inertial observer with field modes in Rindler's region $IV$ (Alice-AntiRob $A\bar R$) and 3) The entanglement between modes in regions $I$ and $IV$ of the Rindler space-time (Rob-AntiRob $R\bar R$). Partitions $AR$ and $A\bar R$ are especially important as these are the partitions in which classical communication is allowed (we will refer to them as CCA bipartitions from now on).

In previous works some aspects about correlations comportment in the presence of horizons were disclosed. These works accounted for the radical differences between fermionic and bosonic entanglement behaviour in the presence of Rindler and event horizons, showing that the real cause of these differences is fermionic/bosonic statistics. This contradicts the naive argument that the differences come from the finite dimensional nature of the fermionic Hilbert space for each frequency as opposed to the built-in infinite dimension of the Fock space for bosons.

However, in previous works it was pointed out that there exist effects of the dimension of the Fock space on the correlations in the bipartition $R\bar R$. As it was shown in \cite{Edu4} these effects clearly distinguish the case of fermions with a different Fock space dimension (Dirac vs Grassman scalar). It was an open question, however, to account for the precise effect of the dimension of the Fock space for each mode in the correlations behaviour of bosonic fields. In this work we will see that, through these effects, we can reach a deeper understanding about the behaviour of correlations in the presence of horizons. 

Nevertheless, in earlier studies it was pointed out that there exist effects of the dimension of the Fock space on the correlations in the bipartition $R\bar R$. As it was shown in \cite{Edu4} these effects clearly distinguish the case of fermions with a different Fock space dimension (Dirac vs Grassman scalar). However, It is still an open question to account for the precise effect of the dimension of the Fock space for each mode on the correlations behaviour of bosonic fields. In this work we will see that analysing these effects we can reach a deeper understanding about the comportment of correlations in the presence of horizons. 

In an attempt to account for the dimensional effects mentioned above, we are going to study a finite dimensional analog to bosonic fields with a limited dimension of each frequency Hilbert space. This means engineering a method to impose a maximum occupation number $N$ in each scalar field frequency mode. The construction of a finite dimensional scalar field state for a non-inertial observer can in itself be problematic, thus an issue which will need to be tackled in order to conduct the proposed analysis.

We will present results about entanglement of the CCA bipartitions that will strengthen the argument discussed in previous works \cite{Edu2,Edu3,Edu4} on the capital importance of statistics in the phenomenon of Unruh decoherence and its role in the black hole information paradox. Specifically we will prove that the behaviour is fundamentally independent of the Fock space dimension. However, bosonic entanglement for $AR$ is slightly sensitive to Fock space dimension variation, in opposition to what happens with fermions. 

We shall point out that those variations are of most interest themselves. Namely they again strongly oppose what is said in previous literature that Unruh decoherence degrades the entanglement quicker as the dimension of the Hilbert space is higher. Instead, we will show that quantum correlations can be more or less quickly degraded for different dimensions depending on the value of the acceleration. This completely banishes the former argument. 

We shall point out that those variations strongly oppose once again what is said in previous literature in which it is argued that the Unruh decoherence degrades the entanglement quicker as the dimension of the Hilbert space is higher. Instead, we will show that quantum correlations can be more or less quickly degraded for different dimensions depending on the value of the acceleration. This completely banishes the former argument. 

We will also show how classical correlations between $AR$ and $A\bar R$ are affected by the bound on the occupation number. Specifically we will show that the effect of imposing a finite dimensional Fock space affects the conservation law for mutual information found in previous works. We will compare this with the fermionic cases and will prove some results about mutual information to be completely universal.

Furthermore, we will show remarkable results concerning correlations between modes in Rindler regions $I$ and $IV$. We will see how they are ruled by bothÊstatistics and Hilbert space dimension. There are differences and similarities between fermions and bosons concerning correlations $R\bar R$. We will analyse the different bosonic cases comparing them with their Fock space dimension fermionic analogs, in order to comprehend the relative importance of dimensionality and statistics in the behaviour of such correlations.

This work is structured as follows. Section \ref{sec2} provides a brief introduction about Rindler space-time and the perspective of an accelerated observer. In section \ref{sec3} we will construct bosonic bipartite entangled states with bounded occupation number. In section \ref{sec4} we will analyse entanglement (subsection \ref{negatsec}) and mutual information (subsection \ref{mutualsec}) for all the possible bipartitions of the system and different values of the occupation number bound $N$, comparing the results of finite dimensional bosons with the fermionic analogs. Finally, we present our conclusions in section \ref{conclusions}. 

\section{Scalar fields as seen from an accelerated observer}\label{sec2}

A uniformly accelerated observer describes the world by means of a set of Rindler coordinates \cite{gravitation}. We need, however, two different set of coordinates in order to cover the whole Minkowski space-time. These sets of coordinates define two causally disconnected regions in Rindler space-time. If we consider that the uniform acceleration $a$ lies on the $z$ axis, the new Rindler coordinates $(t,x,y,z)$ as a function of Minkowski coordinates $(\tilde t,\tilde x,\tilde y,\tilde z)$ are
\begin{equation}\label{Rindlcoordreg1}
a\tilde t=e^{az}\sinh(at),\; a\tilde z=e^{az}\cosh(at),\; \tilde x= x,\; \tilde y= y
\end{equation}
for region I, and
\begin{equation}\label{Rindlcoordreg2}
a\tilde t=-e^{az}\sinh(at),\; a\tilde z=-e^{az}\cosh(at),\; \tilde x= x,\; \tilde y= y
\end{equation}
for region IV.
\begin{figure}\label{fig1}
\includegraphics[width=.45\textwidth]{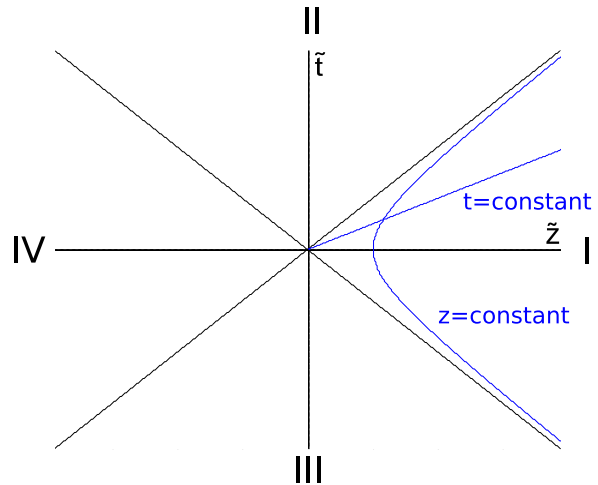}
\caption{Rindler space-time diagram: lines of constant position $z=\text{const.}$ are hyperbolas and all the curves of constant proper time $t$ for the accelerated observer are straight lines that come from the origin. An uniformly accelerated observer Rob travels along a hyperbola constrained to region I}
\end{figure}
As we can see from fig. 1, although we have covered the whole Minkowski space-time with these sets of coordinates, there are two more regions labeled II and III. To map them we would need to trade $\cosh$ by $\sinh$ in equations \eqref{Rindlcoordreg1},\eqref{Rindlcoordreg2}. In these regions, $t$ is a spacelike coordinate and $z$ is a timelike coordinate. However,  considering such regions is not required to describe fields from an accelerated observer perspective \cite{Birrell,gravitation,Alicefalls,AlsingSchul,Edu2}.

The Rindler coordinates $z,t$ go from $-\infty$ to $\infty$ independently in regions I and IV. It means that each region admits a separate quantisation procedure with their corresponding positive and negative energy solutions of Klein-Gordon equations $\psi^{I+}_{k},\psi^{I-}_{k}$ and $\psi^{IV+}_{k},\psi^{IV-}_{k}$. Positive and negative energy solutions will be classified with respect to the future-directed timelike Killing vector in each region. In region I the future-directed Killing vector is
\begin{equation}\label{KillingI}
\partial_t^I=\frac{\partial \tilde t}{\partial t}\partial_{\tilde t}+\frac{\partial\tilde z}{\partial t}\partial_{\tilde z}=a(\tilde z\partial_{\tilde t}+\tilde t\partial_{\tilde z}),
\end{equation}
whereas in region IV the future-directed Killing vector is $\partial_t^{IV}=-\partial_t^{I}$.

This means that solutions in region I, having time dependence $\psi_k^{I+}\sim e^{-ik_0t}$ with $k_0>0$, represent positive energy solutions, whereas solutions in region IV, having time dependence $\psi_k^{I+}\sim e^{-ik_0t}$ with $k_0>0$, are actually negative energy solutions since $\partial^{IV}_t$ points to the opposite direction of $\partial_{\tilde t}$. As I and IV are causally disconnected $\psi^{IV\pm}_{k}$ and $\psi^{I\pm}_{k}$ only have support in their own regions, vanishing outside them.

Let us denote $a_{I,k}^{\phantom{\dagger}},a^{\dagger}_{I,k}$ as the particle annihilation and creation operators in region $I$ and $a^{\phantom{\dagger}}_{IV,k},a^{\dagger}_{IV,k}$ as the particle/antiparticle operators in region IV.

The bosonic operators satisfy the commutation relations $[a^{\phantom{\dagger}}_{\text{R},k},a^\dagger_{\text{R}',k'}]=\delta_{\text{R}\text{R}'}\delta_{kk'}$. The subscript R notates the Rindler region of the operator $\text{R}=\{I,IV\}$.  All other commutators  are zero.

We can relate Minkowski and Rindler creation and annihilation operators by taking appropriate inner products and computing the so-called Bogoliubov coefficients \cite{Takagi,Jauregui,Birrell,AlsingSchul}. This allows us to translate Minkowskian states into Rindler coordinates as we will present in the next section. For a scalar field, the Bogoliubov relationships for the annihilation operator of modes with positive frequency are
\begin{eqnarray}\label{bogoboson}
 a_{M,k}&=&\cosh r\, a_{I,k} - \sinh r\, a^\dagger_{IV,-k}
\end{eqnarray}
where
\begin{equation}\label{defr1}
\tanh r=e^{-\pi \frac{k_0c}{a}}
\end{equation}

\section{Limiting the occupation number}\label{sec3}

The Minkowski vacuum state for the scalar field is defined by the tensor product of each frequency mode vacuum
\begin{equation}\label{vacuas}\ket0=\bigotimes_{k}\ket{0_k}\end{equation}
such that it is annihilated by $a_{k}$ for all values of $k$.

Through this work we will operate under the single mode approximation (SMA) \cite{Alsingtelep,AlsingMcmhMil}. In any case in \cite{Edu3} we discussed the implications of such approximation and, for the purposes of this work, going beyond SMA will not give relevant differences.

The vacuum state for a $k$-momentum mode of a scalar field seen from the perspective of an accelerated observer is
\begin{equation}\label{scavacinf}
\ket{0}_M=\frac{1}{\cosh r}\sum_{n=0}^\infty \tanh^n r \ket{n}_I\ket{n}_{IV}
\end{equation}
Notice that here and from now on we are dropping the $k$ label as we are working under SMA.

The Minkowskian one particle state results from applying the creation operator to the vacuum state. Its translation to the Rindler basis is
\begin{equation}\label{unoinf}
\ket{1}_M=\frac{1}{\cosh^2 r}\sum_{n=0}^{\infty} \tanh^n r \,\sqrt{n+1}\ket{n+1}_I\ket{n}_{IV}
\end{equation} 	
Now we will consider the following maximally entangled state in the Minkowsky basis
\begin{equation}
\label{entangledscainf}\ket{\Psi}=\frac{1}{\sqrt{2}}\left(\ket{0}_M\ket{0}_M+\ket{1}_M\ket{1}_M\right)
\end{equation}
This is a qubit state which is a superposition of the bipartite vacuum and the bipartite one particle state. 

For our purposes we need to limit the dimension of the Hilbert space. To do so we are going to limit the maximum occupation number for the Rindler modes up to $N$.

As a first approach, one could consider bosons with a limited dimension of the Hilbert space by imposing occupation number bounds on the states \eqref{scavacinf}, \eqref{unoinf} to model impenetrable bosons as done to illustrate some qualitative arguments in \cite{Edu4}. 

However, if we want to go beyond the qualitative effects and do a completely rigorous analysis we would run into important problems. Namely, we would not be able to normalise both states  \eqref{scavacinf}, \eqref{unoinf} simultaneously. In other words, the translation into Rindler coordinates of the Minkowskian creation operator applied on the vacuum state do not preserve normalisation and it is no longer true that applying the annihilation operator to the one particle state we recover the vacuum of the theory. Therefore the second quantisation rules of bosonic fields would be ill-defined (for instance, problems would appear when applying the commutator to the one particle state). As statistics is fundamental to explain the Unruh decoherence mechanism, a rigorous analysis would require that we consider the vacuum of our theory as expressed in \eqref{scavacinf} with an unbounded occupation number.

Alternately, we will define finite dimension analogs to the vacuum and one particle states
\begin{equation}\label{scavac}
\ket{0_N}_M=\frac{1}{\cosh r}\sum_{n=0}^N \tanh^n r \ket{n}_I\ket{n}_{IV}
\end{equation}
\begin{equation}\label{uno}
\ket{1_N}_M=\frac{1}{\cosh^2 r}\sum_{n=0}^{N-1} \tanh^n r \,\sqrt{n+1}\ket{n+1}_I\ket{n}_{IV}
\end{equation} 	
in which we have cut off the higher occupation numbers and thus these two states are not exactly the vacuum of our theory and the first excitation. Instead, they could be understood as approximations in which Rob is not able to notice occupation numbers larger than $N$. This simple construct allows us to consider a bounded occupation number along with bosonic statistics. Therefore we can now disentangle the statistical effects from the ones derived from the dimensionality of the Hilbert space. This limiting our states to a subset of relevant modes is a similar procedure to the one used to carry out the SMA \cite{Alsingtelep,AlsingMcmhMil}.

We will then consider the following entangled state in Minkowsky coordinates
\begin{equation}
\label{entangledsca}\ket{\Psi}=\frac{1}{C_N(r)}\left(\ket{0}_M\ket{0_N}_M+\ket{1}_M\ket{1_N}_M\right)
\end{equation}
in which the one particle state and the vacuum for Rob are substituted by the bounded occupation number approximations.

Notice that a factor $1/C_N(r)$ must now be included as our occupation number cutoff implies that $\ket{0_N}_M$ and $\ket{1_N}_M$ are not normalised. Its value is
\begin{equation}
C_N(r)=\sqrt{\braket{0_N}{0_N}_M+\braket{1_N}{1_N}_M}
\end{equation}
explicitly
\begin{equation}
C_N(r)=\sqrt{2-\tanh^{2N}r\left(\tanh^2 r + 1 + \frac{N}{\cosh^2r}\right)}
\end{equation}
In the limit $N\rightarrow\infty$, $C_{N}(r)\rightarrow\sqrt2$ recovering the standard scalar maximally entangled state \eqref{entangledscainf}.

This normalisation will also guarantee that Unruh decoherence will not affect higher occupation number modes, in similar way as SMA restricts the decoherence to a single frequency.

The density matrix for the whole tripartite state, which includes modes in both sides of the horizon along with Minkowskian modes, is built from \eqref{entangledsca} changing to Rindler coordinates for Rob
\begin{equation}\label{tripasca}
\rho^{AR\bar R}=\proj{\Psi}{\Psi}
\end{equation}

\begin{eqnarray}
\label{AR1}\rho^{AR}&=&\tr_{IV}\rho^{AR\bar R}\\*
\label{AAR1}\rho^{A\bar R}&=&\tr_{I}\rho^{AR\bar R}\\*
\label{RAR1}\rho^{R\bar R}&=&\tr_{M}\rho^{AR\bar R}
\end{eqnarray}
and the density matrix for each individual subsystem
 \begin{eqnarray}
\label{A1}\rho^{A}&=&\tr_{I}\rho^{AR}=\tr_{IV}\rho^{A\bar R}\\*
\label{R1}\rho^{R}&=&\tr_{IV}\rho^{R\bar R}=\tr_{M}\rho^{AR}\\*
\label{aR1}\rho^{\bar R}&=&\tr_{I}\rho^{R\bar R}=\tr_{M}\rho^{A \bar R}
\end{eqnarray}

The bipartite systems are characterized by the following density matrices
\begin{eqnarray}\label{rhoars1}
\rho^{AR}&=&\!\left\{\sum_{n=0}^{N-1}\frac{\tanh^{2n}r}{\cosh^2r}\left[\proj{0n}{0n}+\frac{\sqrt{n+1}}{\cosh r}\Big(\proj{0n}{1\, n+1}\right.\nonumber\right.\\*
&&\left.+\proj{1\, n+1}{0n}\Big)+\frac{n+1}{\cosh^2 r}\proj{1\,n+1}{1\,n+1}\right]+\nonumber\\*
&&\left.+\frac{\tanh^{2N}r}{\cosh^2r}\proj{0N}{0N}\right\}\frac{1}{C_N(r)^2}
\end{eqnarray}
\begin{eqnarray}\label{rhoa-rs1}
\nonumber\rho^{A\bar R}&\!\!\!=\!\!\!&\left\{\sum_{n=0}^{N-1}\frac{\tanh^{2n}r}{\cosh^2r}\left[\proj{0n}{0n}\!+\!\frac{\sqrt{n+1}}{\cosh r}\tanh r\Big(\!\ket{0\,n+1}\right.\right.\\*
&&\!\!\!\!\!\!\left.\times\bra{1 n}+\proj{1 n}{0\,n+1}\Big)+\frac{n+1}{\cosh^2 r}\proj{1n}{1n}\right]+\nonumber\\*
&& \!\!\!\!\!\!\left.+\frac{\tanh^{2N}r}{\cosh^2r}\proj{0N}{0N}\right\}\frac{1}{C_N(r)^2}
\end{eqnarray}
\begin{eqnarray}\label{rhor-rs1}
\nonumber\rho^{R\bar R}\!\!&=&\Bigg\{\sum_{\substack{n=0\\m=0}}^{N}\frac{\tanh^{n+m}r}{\cosh^2r}\proj{nn}{mm}+\sum_{\substack{n=0\\m=0}}^{N-1}\frac{\tanh^{n+m}r}{\cosh^4r} \\*
&&\!\!\!\!\!\!\!\!\!\!\!\times\sqrt{n+1}\sqrt{m+1}\proj{n+1\,n}{m+1\,m}\Bigg\}\frac{1}{C_N(r)^2}
\end{eqnarray}
where the bases are respectively
\begin{eqnarray}\label{barbolbasis}
 \ket{nm}&=&\ket{n^A}_{M}\ket{m^R}_{I}\\*
\ket{nm}&=&\ket{n^A}_{M}|m^{\bar R}\rangle_{IV}\\*
\ket{nm}&=&\ket{n^R}_{I}|m^{\bar R}\rangle_{IV}
\end{eqnarray}
for \eqref{rhoars1}, \eqref{rhoa-rs1} and \eqref{rhor-rs1}.

On the other hand, the density matrices for the individual subsystems \eqref{A1}, \eqref{R1},\eqref{aR1} are
\begin{equation}\label{Robpartial}
\rho^{R}=\frac{1}{C_N(r)^2}\sum_{n=0}^N\frac{\tanh^{2n}r}{\cosh^2r}\left[1+\frac{n}{\sinh^2r}\right]\proj{n}{n}
\end{equation}
\begin{eqnarray}\label{ARobpartial}
\nonumber\rho^{\bar R}&=&\frac{1}{C_N(r)^2}\left[\sum_{n=0}^{N-1}\frac{\tanh^{2n}r}{\cosh^2r}\left(1+\frac{n+1}{\cosh^2r}\right)\proj{n}{n}\right.\\*
&&\left.+\frac{\tanh^{2N}r}{\cosh^2r}\proj{N}{N}\right]
\end{eqnarray}
\begin{equation}\label{AlicedeAliceRob}
\rho^{A}=\frac{1}{C_N(r)^2}\left(D^0_N(r)\proj{0}{0}+D^1_N(r)\proj{1}{1}\right)
\end{equation}
Where
\begin{equation}\label{S1}
D^0_N(r)=\sum_{n=0}^N\frac{\tanh^{2n}r}{\cosh^2 r}=1-(\tanh r)^{2(N+1)}
\end{equation}
\begin{equation}\label{S2}
D^1_N(r)\!=\!\! \sum_{n=0}^{N-1}(n+1)\frac{\tanh^{2n}r}{\cosh^2 r} =1-\left(1+\dfrac{N}{\cosh^2 r}\right) \tanh^{2 N}\!r
\end{equation}
Notice that $D^0_N(r)+D^1_N(r)=C_N^2(r)$ and consequently all the density matrix traces are 1 as it must be. As all the probability is within the modes that we are considering, all the possible decoherence is confined to the finite occupation number Hilbert space we are studying.

As an effect of the imposition of the finite dimension $\rho_A\rightarrow\proj00$ as $a\rightarrow\infty$ for any finite $N$, but it tends to $\frac12(\proj00+\proj11)$ when $N\rightarrow\infty$ for all $a$. It is important to notice that both limits do not commute. The limit $N\rightarrow\infty$ should be taken first in order to recover the standard scalar field result.

\section{Analysis of correlations}\label{sec4}

In this section we will analyse the correlations tradeoff among all the possible bipartitions of the system. We will account for the entanglement by means of the negativity, and we will study the total correlations by means of the mutual information, which accounts for both classical and quantum.

\subsection{Quantum correlations}\label{negatsec}

Negativity is an entanglement monotone defined as the sum of the negative eigenvalues of the partial transpose density matrix for the system, which is defined as the transpose of only one of the subsystem q-dits in the bipartite density matrix. If $\sigma_i$ are the eigenvalues of $\rho^{pT}_{AB}$ then
\begin{equation}\label{negativitydef}
\mathcal{N}_{AB}=\frac12\sum_{ i}(|\sigma_i|-\sigma_i)=-\sum_{\sigma_i<0}\sigma_i
\end{equation}

To compute the negativity we need the partial transpose of the bipartite density matrices \eqref{rhoars1}, \eqref{rhoa-rs1} and \eqref{rhor-rs1}, which we will notate as $\eta^{AR}$, $\eta^{A\bar R}$ and $\eta^{R \bar R}$  respectively.

\begin{eqnarray}\label{etaARs}
\eta^{AR}\!\!&=&\!\left\{\sum_{n=0}^{N-1}\frac{\tanh^{2n}r}{\cosh^2r}\left[\proj{0n}{0n}+\frac{\sqrt{n+1}}{\cosh r}\Big(\proj{0\, n+1}{1n}\right.\nonumber\right.\\*
&&\left.+\proj{1n}{0\, n+1}\Big)+\frac{n+1}{\cosh^2 r}\proj{1\,n+1}{1\,n+1}\right]+\nonumber\\*
&&\left.+\frac{\tanh^{2N}r}{\cosh^2r}\proj{0N}{0N}\right\}\frac{1}{C_N(r)^2}
\end{eqnarray}
\begin{eqnarray}\label{etaAaRs}
\nonumber\eta^{A\bar R}&\!\!\!=\!\!\!&\left\{\sum_{n=0}^{N-1}\frac{\tanh^{2n}r}{\cosh^2r}\left[\proj{0n}{0n}\!+\!\frac{\sqrt{n+1}}{\cosh r}\tanh r\Big(\!\ket{0n}\right.\right.\\*
&&\!\!\!\!\!\!\left.\times\bra{1 \,n+1}+\proj{1\, n+1}{0n}\Big)+\frac{n+1}{\cosh^2 r}\proj{1n}{1n}\right]+\nonumber\\*
&& \!\!\!\!\!\!\left.+\frac{\tanh^{2N}r}{\cosh^2r }\proj{0N}{0N}\right\}\frac{1}{C_N(r)^2}
\end{eqnarray}
\begin{eqnarray}\label{etaRaRs}
\nonumber\eta^{R\bar R}&=&\Bigg\{\sum_{\substack{n=0\\m=0}}^{N}\frac{\tanh^{n+m}r}{\cosh^2r}\proj{nm}{mn}+\sum_{\substack{n=0\\m=0}}^{N-1}\frac{\tanh^{n+m}r}{\cosh^4r} \\*
&&\!\!\!\!\!\!\!\!\!\!\!\times\sqrt{n+1}\sqrt{m+1}\proj{n+1\,m}{m+1\,n}\Bigg\}\frac{1}{C_N(r)^2}
\end{eqnarray}

In the following subsections we shall compute the negativity of each bipartition of the system.
 
\subsubsection{Bipartition Alice-Rob}

Apart from the diagonal elements corresponding to $\proj{00}{00}$ and $\proj{1N}{1N}$ (which form two $1\times1$ blocks themselves), the partial transpose of the density matrix $\rho^{A R} $ \eqref{etaARs} has a $2\times2$ block structure in the basis $\{ \ket{0\, n+1},\ket{1 n}\}_{n=0}^{N-1}$
\begin{equation}\label{blocks}
\frac{\tanh^{2n}r}{C(r)^2\cosh^2 r}
\left(\!\begin{array}{cc}
\tanh^2 r & \dfrac{\sqrt{n+1}}{\cosh r}\\
\dfrac{\sqrt{n+1}}{\cosh r} & \dfrac{n}{\sinh^2r}
\end{array}\!\right)
\end{equation}
Hence, the eigenvalues of \eqref{etaARs} are
\begin{eqnarray}
\nonumber\lambda^\pm_n&=&\frac{\tanh^{2n} r}{2C_N(r)^2\cosh^2 r}\Bigg[\left(\frac{n}{\sinh^2r}+\tanh^2 r\right)\\*
&&\nonumber\left.\pm\sqrt{\left(\frac{n}{\sinh^2r}+\tanh^2 r\right)^2+\frac{4}{\cosh^2 r}}\right]_{n=0}^{N-1}\\*
\nonumber\lambda_N&=&\frac{1}{C_N(r)^2\cosh^2r};\qquad \lambda_{N+1}=\frac{N(\tanh r)^{2N-2}}{C_N(r)^2\cosh^4 r}\\*
\end{eqnarray}
Where the notation $\displaystyle{|_{n=a_1}^{a_N}}$ means that $n$ takes all the integer values from $a_1$ to $a_N$.

Therefore the negativity for this bipartition is
\begin{eqnarray}
\nonumber\mathcal{N}^{AR}&=&\sum_{n=0}^{N-1}\frac{\tanh^{2n} r}{2C_N(r)^2\cosh^2 r}\left|\left(\frac{n}{\sinh^2r}+\tanh^2 r\right)\right.\\*
&&\!\!\!\!\!\!\!\left.-\sqrt{\left(\frac{n}{\sinh^2r}+\tanh^2 r\right)^2+\frac{4}{\cosh^2 r}}\right|
\end{eqnarray}

\begin{figure}[h]
\includegraphics[width=.50\textwidth]{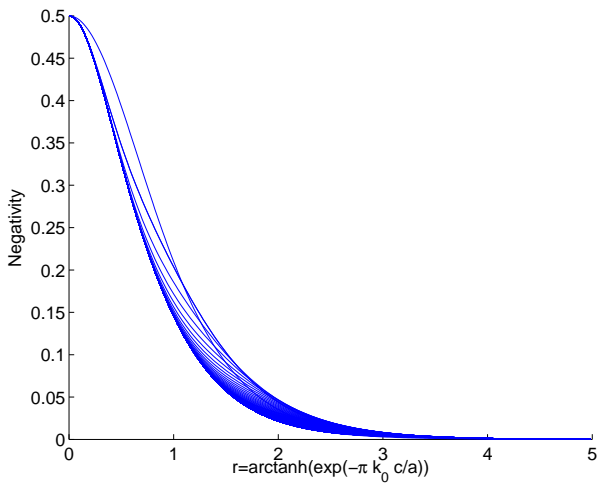}
\caption{(Colour online) Bundle of curves displaying the negativity for the bipartition $AR$ for all values of $N$ as a function of the acceleration}
\label{bundle}
\end{figure}

Figure \ref{bundle} shows the behaviour of negativity for all values of $N$, which is clearly similar for all cases no matter how many dimensions we are allowing for each mode. Despite this, negativity $AR$ is slightly sensitive to dimension variations. This is a difference with the fermionic cases, in which varying the dimension does not affect $\mathcal{N}_\text{fermionic}^{AR}$  at all in  \cite{Edu3}.

\begin{figure}[h]
\includegraphics[width=.50\textwidth]{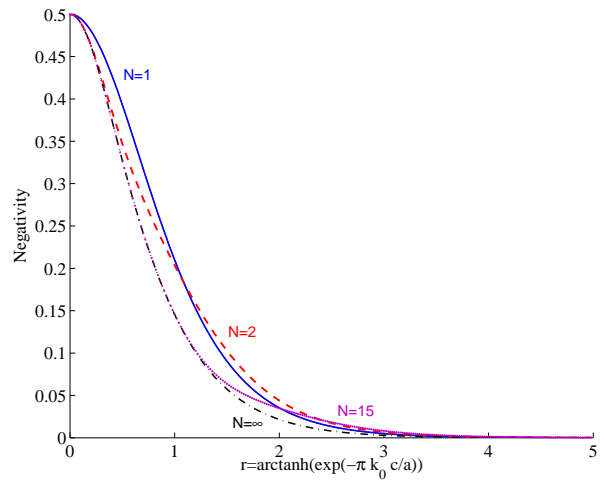}
\caption{(Colour online): Same as fig. \ref{bundle} for selected choices of $N$ showing the existence of crossing points. To the right (left) of these points entanglement degrades less (more) for lesser $N$. Solid blue line $N=1$, dashed red line $N=2$, dotted purple line $N=15$, black dash-dotted line $N=\infty$.}
\label{negaARcomp}
\end{figure}

$\mathcal{N}^{AR}$ is shown in figure \ref{negaARcomp} as a function of $r$ for different values of $N$, comparing them with the case $N=1$. 
A very interesting result emerges here, for any pair of values for the maximum occupation number $N_1<N_2$ both negativity curves cross in a point $a=a_c(N_1,N_2)$. This means that for any finite value of the Hilbert space dimension there is a region $a<a_c(N_1,N_2)$ (low accelerations) in which entanglement is more degraded for higher dimension, and another region $a>a_c(N_1,N_2)$ (high accelerations)  in which entanglement is more degraded for lower dimension.

\begin{figure}[h]
\includegraphics[width=.50\textwidth]{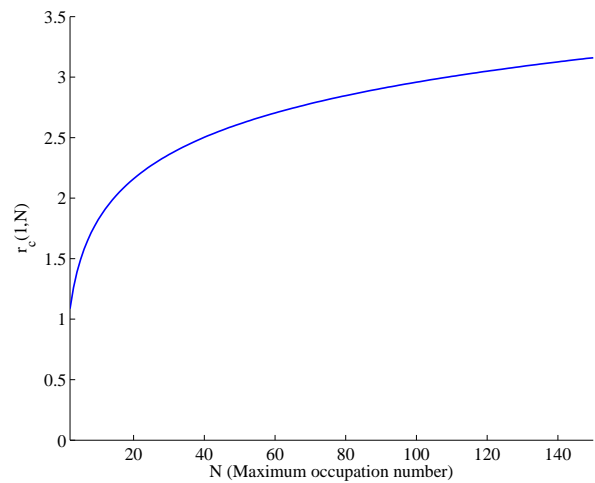}
\caption{(Colour online) Values of $r$ in which negativity curves for occupation number $N$ cross the negativity curve for $N=1$. In the region above the curve entanglement degrades faster for $N=1$ than for $N>1$.}
\label{rcritical}
\end{figure}

This disagrees the naive argument that higher dimension would lead to higher Unruh decoherence which is not necessarily true. Figure \ref{rcritical} shows the behaviour of $r_c(1,N)$ as $N$ grows. The crossing point with the negativity curve for $N=1$ grows as we consider larger $N$ curves. $r_c(N_1,N_2)$ is related with $a_c(N_1,N_2)$ by means of the relationship \eqref{defr1}.

For the limit $N_2\rightarrow\infty$,  $a_c(N_1,N_2)\rightarrow\infty$, which means that infinite dimension negativity is below all the finite dimensional curves.
\subsubsection{Bipartition Alice-AntiRob}

Excepting the diagonal elements corresponding to $\proj{10}{10}$ and $\proj{0N}{0N}$  (which form two $1\times1$ blocks themselves), the partial transpose of the density matrix $\rho^{A \bar R}$ \eqref{etaAaRs} has a $2\times2$ block structure in the basis $\{ \ket{0 n},\ket{1\, n+1}\}_{n=0}^{N-1}$ 
\begin{equation}\label{blocksbosAaR}
\frac{\tanh^{2n}r}{C(r)^2\cosh^2 r}
\left(\!\begin{array}{cc}
1 & \dfrac{\tanh r}{\cosh r}\sqrt{n+1}\\[3mm]
\dfrac{\tanh r}{\cosh r}\sqrt{n+1} & \dfrac{\tanh^2 r}{\cosh^2r}(n+2)
\end{array}\!\right)
\end{equation}
Hence, the eigenvalues of \eqref{etaAaRs} are
\begin{eqnarray}
\nonumber\lambda^{\pm}_n&=&\frac{\tanh^{2n}r}{2C(r)^2\cosh^2 r}\Bigg[\left(1+(n+2)\frac{\tanh^2 r}{\cosh^2 r}\right)\\*
&&\!\!\!\!\left.\pm\sqrt{\left(1+(n+2)\frac{\tanh^2 r}{\cosh^2 r}\right)^2-\frac{4\tanh^2 r}{\cosh^2 r}}\right]_{n=0}^{N-1}\nonumber\\*
\lambda_N&=&\frac{1}{C(r)^2\cosh^4 r};\qquad \lambda_{N+1}=\frac{\tanh^{2N}r}{C(r)^2\cosh^2 r}
\end{eqnarray}
Therefore, the negativity for this bipartition is always $0$, independently of the value of the acceleration parameter and the occupation number bound $N$.

From this results can be concluded that limiting the dimension has no effect in the creation or not of quantum correlations between Alice and AntiRob. As far as the field is bosonic, no entanglement is created in the CCA bipartitions of the system no matter how we limit the dimension of the Hilbert space.

\subsubsection{Bipartition Rob-AntiRob}

The partial transpose of the density matrix $\rho^{R\bar R}$  \eqref{etaRaRs} has a block structure. Namely, it is formed by $2N+1$ blocks whose dimension varies. In the following we will detailedly analyse the blocks.

\begin{enumerate}
\item First of all, we have $N+1$ blocks $\left\{M_D\right\}_{D=1}^{N+1}$  which are  endomorphisms that act in the subspace (of dimension $D$) expanded by the basis $B_{D}=\{\ket{mn}\}$ in which $m+n=D-1\le N$. 
\item Then we have $N$ more blocks $\left\{M'_D\right\}_{D=1}^N$ that  act in the subspace (of dimension $D$) expanded by the basis $B'_{D}=\{\ket{m'n'}\}$ in which \mbox{$m'+n'=2N-D+1>N$}. Notice that not all the possible $m'$ and $n'$ are allowed due to the limitation to the occupation number $m',n'\le N$.
\end{enumerate}

As an example which will perfectly clarify this construction, if $N=4$ there will be 9 blocks, $M_1,M_2,M_3,M_4,M_5, M'_4,M'_3,M'_2,M'_1$ each one is an endomorphism which acts in the subspace expanded by the bases
\begin{eqnarray}
\nonumber B_1&=&\left\{\ket{00}\right\}\\*
\nonumber B_2&=&\left\{\ket{01},\ket{10}\right\}\\*
\nonumber B_3&=&\left\{\ket{02},\ket{20},\ket{11}\right\}\\*
\nonumber B_4&=&\left\{\ket{03},\ket{30},\ket{12},\ket{21}\right\}\\*
\nonumber B_5&=&\left\{\ket{04},\ket{40},\ket{13},\ket{31},\ket{22}\right\}\\*
\nonumber B'_4&=&\left\{\ket{14},\ket{41},\ket{23},\ket{32}\right\}\\*
\nonumber B'_3&=&\left\{\ket{24},\ket{42},\ket{33}\right\}\\*
\nonumber B'_2&=&\left\{\ket{34},\ket{43}\right\}\\*
\nonumber B'_1&=&\left\{\ket{44}\right\}\\*
\end{eqnarray}
respectively.

In this fashion, the whole matrix is an endomorphism within the subspace R=$\bigoplus_{i=1}^{N+1} S_i\oplus\bigoplus_{j=1}^N S'_j$, being $S_i$ the subspace (of dimension $D=i$) expanded by the basis $B_i$ and $S'_j$ the subspace (of dimension $D=j$) expanded by the basis $B'_j$ .

The blocks $M_1,\dots,M_{N+1}$ and $M'_1,\dots,M'_{N}$ which form the matrix \eqref{etaRaRs} have the following form
\begin{equation}\label{blockss}
M_D=\left(\!
\begin{array}{cccccccc}
0  & a_1  & 0 & 0 & \cdots & \cdots& \cdots& 0 \\
a_1 & 0 & a_2 & 0 & \cdots & \cdots& \cdots & 0\\
0 & a_2 & 0 & a_3 & \cdots & \cdots& \cdots& 0\\
0 & 0 & a_3 &0 & a_4 & \cdots& \cdots& 0\\
0 & 0 & 0 &  \ddots &\ddots &  \ddots &\cdots& 0\\
\vdots  & \vdots  & \vdots  & \vdots  & \ddots  & \ddots  & \ddots  & \vdots \\
0 & 0 & 0 &  0 &\cdots&  \ddots &0& a_{D-1}\\
0 & 0 & 0 &  0 &0&  \dots &a_{D-1}& a_{D}\\
\end{array}\!\right)
\end{equation}
\begin{equation}
M'_D=\left(\!
\begin{array}{cccccccc}
0  & b_{1}  & 0 & 0 & \cdots & \cdots& \cdots& 0 \\
b_{1} & 0 & b_{2} & 0 & \cdots & \cdots& \cdots & 0\\
0 & b_{2} & 0 & b_{3} & \cdots & \cdots& \cdots& 0\\
0 & 0 & b_{3} &0 & b_{4} & \cdots& \cdots& 0\\
0 & 0 & 0 &  \ddots &\ddots &  \ddots &\cdots& 0\\
\vdots  & \vdots  & \vdots  & \vdots  & \ddots  & \ddots  & \ddots  & \vdots \\
0 & 0 & 0 &  0 &\cdots&  \ddots &0& b_{D-1}\\
0 & 0 & 0 &  0 &0&  \dots &b_{D-1}& b_{D}\\
\end{array}\!\right)
\end{equation}

The matrix elements $a_n$ and $b_n$ are defined as follows
\begin{eqnarray}
\nonumber a_{2l+1}&=&\frac{(\tanh r)^{D-1}}{C(r)^2\cosh^2 r}\\*
\nonumber a_{2l}&=&\sqrt{D-l}\,\sqrt{l}\frac{(\tanh r)^{D-2}}{C(r)^2\cosh^4r}\\*
\nonumber b_{2l+1}&=&\frac{(\tanh r)^{2N-D+1}}{C(r)^2\cosh^2 r}\\*
\nonumber b_{2l}&=&\sqrt{N+1-l}\,\sqrt{l+N-D+1}\frac{(\tanh r)^{2N-D}}{C(r)^2\cosh^4r}\\*
\end{eqnarray}

Notice that the elements are completely different when the value of the label $n$ is odd or even.

\begin{figure}[h]
\includegraphics[width=.50\textwidth]{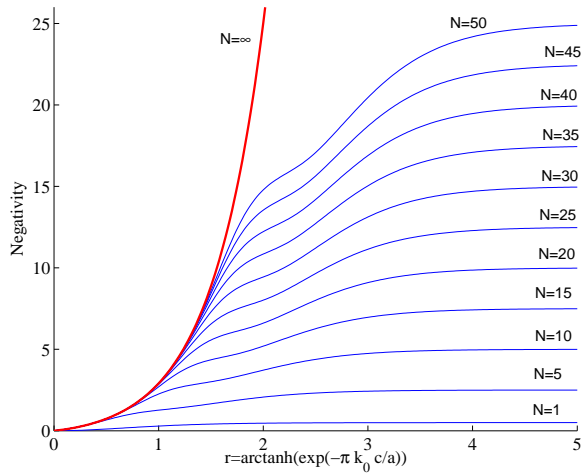}
\caption{(Colour online) Negativity $R\bar R$ for different values of $N$, showing the upper bound reached when $a\rightarrow\infty$. Negativity diverges in the black hole limit only for $N\rightarrow\infty$.}
\label{negaRAR}
\end{figure}

As the whole matrix is the direct sum of the blocks
\begin{equation}
\eta^{R \bar R}=\left(\bigoplus_{D=1}^{N+1} M_D\right) \oplus \left(\bigoplus_{D=1}^{N} M'_D\right)
\end{equation}
the eigenvalues and, specifically, the negative eigenvalues of $\eta^{R \bar R}$ would be the negative eigenvalues of all the blocks $M_D$ and $M'_D$  gathered togheter, which can be easily computed numerically.  Figure \ref{negaRAR} shows the behaviour of $\mathcal{N}^{R\bar R}$ with $r$ and for different values of $N$.

We can now compare the finite $N$ bosonic case with their same dimension analog for fermions. Namely, a Grassman scalar field (spinless fermion) has the same Hilbert space dimension as the scalar case with $N=1$, the only difference is the anticommutation of the field operators instead of the commutation which applies for bosons. On the other hand, scalars limited to $N=4$ and $N=2$ can be considered as two different analogs to the Dirac field as the former has the same Hilbert space dimension as Dirac modes and the latter would share the same possible maximum occupation number. 

\begin{figure}[h]
\includegraphics[width=.50\textwidth]{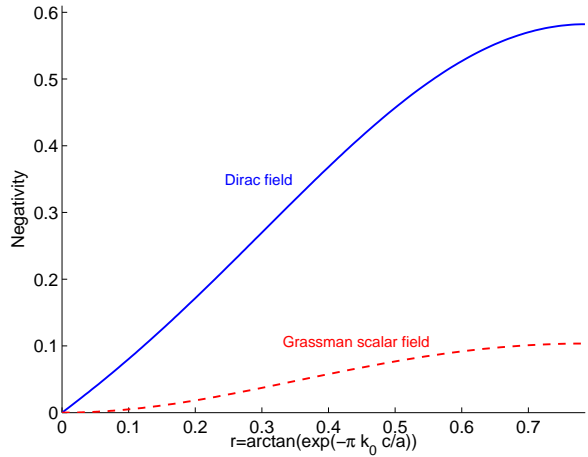}
\caption{(Colour online) Negativity of the bipartition $R\bar R$ for fermion fields. Namely Grassman scalar (red dashed line) and Dirac (blue solid line). Negativity upper bound is greater for the Dirac case as $\dim(\mathcal{H}_{\text{Dirac}})>\dim(\mathcal{H}_{\text{Grassman.}})$ Notice that here $r=\tan(e^{-\pi k_0/a})$ instead of the hyperbolic tangent and therefore $r\rightarrow\pi/4\Rightarrow a\rightarrow\infty$. }
\label{comparison1}
\end{figure}

\begin{figure}[h]
\includegraphics[width=.50\textwidth]{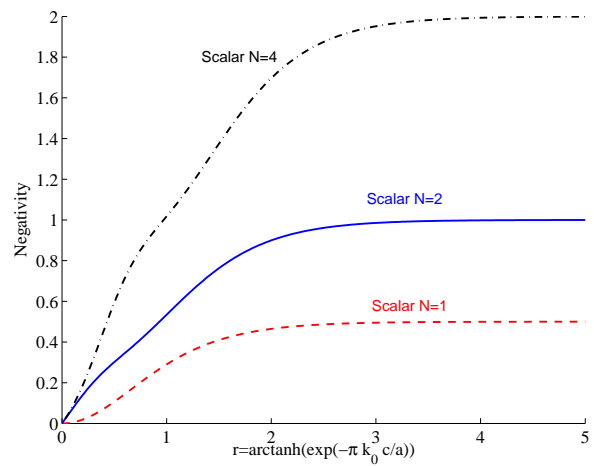}
\caption{(Colour online) Same as fig. \ref{comparison1} but for bounded occupation number scalar fields. $N=1$ (red dashed line) is dimensionally analogous to the Grassman scalar case. $N=4$ (black dash-dotted line) is dimensionally analogous to the Dirac case. $N=2$ (blue solid line) is analogous to the Dirac field in maximum occupation number. Notice that more entanglement is grown for greater $N$.}
\label{comparison2}
\end{figure}

This comparison can be seen in figures \ref{comparison1}, \ref{comparison2}. We see that the comportment is similar (monotonic growth from zero to a finite limit for $a\rightarrow\infty$) but the functional dependence is still very different in both cases. Specifically, as $a$ increases the bosonic cases grow a higher entanglement between the modes of the field on both sides of the horizon than the same dimension fermionic analogs. The ability to create correlations between modes in $I$ and $IV$ is related with the dimension of the Hilbert space but not only. Statistics plays also an important role.

This clearly shows another important difference between fermionic and bosonic fields. Pauli exclusion principle prevents the total degradation of fermionic  entanglement in CCA bipartitions, whereas, conversely, impedes entanglement creation between $R\bar R$.

\subsection{Mutual information}\label{mutualsec}

Mutual information accounts for correlations (both quantum and classical) between two different partitions of a system. It is defined as
\begin{equation}\label{mutualdef}
I_{AB}=S_A+S_B-S_{AB}
\end{equation}
where $S_A$, $S_B$ and $S_{AB}$ are respectively the Von Neumann entropies for the individual subsystems $A$ and $B$ and for the joint system $AB$.

To compute the mutual information  for each bipartition we will need the eigenvalues of the corresponding density matrices. We shall go through all the process detailedly in the lines below.

\subsubsection{Bipartition Alice-Rob}

Excepting the element $\proj{0N}{0N}$ (which forms a $1\times1$ block itself) the density matrix for the system Alice-Rob \eqref{rhoars1} consists on $N$ $2\times2$ blocks in the basis $\{\ket{0 n},\ket{1\, n+1}\}_{n=0}^{N-1}$ which have the form
\begin{equation}
\frac{\tanh^{2n}r}{C_N(r)^2\cosh^2r }
\left(\!\begin{array}{cc}
1 & \dfrac{\sqrt{n+1}}{\cosh r}\\
\dfrac{\sqrt{n+1}}{\cosh r} & \dfrac{n+1}{\cosh^2r}
\end{array}\!\right)
\end{equation}
Hence, the eigenvalues of \eqref{rhoars1} are
\begin{eqnarray}\label{eigAR}
\nonumber\lambda_n&=&\left.\frac{\tanh^{2n}r}{C_N(r)^2\cosh^2r}\left(1+\frac{n+1}{\cosh^2 r}\right)\right|_{n=0}^{N-1}\\*
\lambda_N&=&\frac{\tanh^{2N} r}{C_N(r)^2\cosh^2r}
\end{eqnarray}
Along with $N$ identically zero eigenvalues.

\subsubsection{Bipartition Alice-AntiRob}

Except from the diagonal element corresponding to $\proj{00}{00}$  (which forms one $1\times1$ block itself) the density matrix for the system Alice-AntiRob \eqref{rhoa-rs1} consists on $N-1$ $2\times2$ blocks in the basis $\{\ket{0 n},\ket{1\, n-1}\}_{n=1}^N$ which have the form
\begin{equation}
\frac{\tanh^{2n}r}{C_N(r)^2\cosh^2r}
\left(\!\begin{array}{cc}
1 & \dfrac{\sqrt{n}}{\sinh r} \\
\dfrac{\sqrt{n}}{\sinh r} & \dfrac{n}{\sinh^2 r}\\
\end{array}\!\right)
\end{equation}
Therefore the eigenvalues of \eqref{rhoa-rs1} are
\begin{eqnarray}\label{eigAaR}
\lambda_n&=&\left.\frac{\tanh^{2n}r}{C_N(r)^2\cosh^2r}\left(1+\frac{n}{\sinh^2r}\right)\right|_{n=0}^{N}
\end{eqnarray}
along with $N$ identically zero eigenvalues.

\subsubsection{Bipartition Rob-AntiRob}

The density matrix for Rob-AntiRob  \eqref{rhor-rs1} consists in the direct sum of two blocks 
\begin{equation}\label{newdoe}
\rho^{R\bar R}=X\oplus Y
\end{equation}
of dimensions $\dim(X)=N+1$, $\dim(Y)=N$.
The matrix elements of $X$ and $Y$ are
\begin{equation}
X_{ij}=\frac{(\tanh r)^{i+j-2}}{C_N(r)^2\cosh^2r }\qquad Y_{ij}=\sqrt{i}\sqrt{j} \frac{(\tanh r)^{i+j-2}}{C_N(r)^2\cosh^4r}
\end{equation}
in the bases $\left\{\ket{nn}\right\}_{n=0}^{N}$ and $\left\{\ket{n+1\,n}\right\}_{n=0}^{N-1}$ respectively.

It is easy to see that $\operatorname{rank}(X)=\operatorname{rank}(Y)=1$. This means that all the eigenvalues of  \eqref{rhor-rs1} are zero  except for two of them, which we can readily compute
\begin{equation}\label{eigRaR}
\lambda^{R\bar R}_{X}=\frac{D_N^0(r)}{C(r)^2}\qquad\lambda^{R\bar R}_{Y}= \frac{D_N^1(r)}{C(r)^2}
\end{equation}
where $D_N^0(r)$ and $D_N^1(r)$ are given by \eqref{S1} and \eqref{S2}.

\subsubsection{Von Neumann entropies for each subsystem and mutual information}

To compute the Von Neumann entropies we need the eigenvalues of every bipartition and the individual density matrices. The eigenvalues of $\rho^{AR}$, $\rho^{A\bar R}$, $\rho^{R\bar R}$ are respectively \eqref{eigAR}, \eqref{eigAaR} and \eqref{eigRaR}.

The eigenvalues of the individual systems density matrices can be directly read from \eqref{Robpartial}, \eqref{ARobpartial} and \eqref{AlicedeAliceRob} since $\rho^R$, $\rho^{\bar R}$ and $\rho^A$ have diagonal forms in the Fock basis. The Von Neumann entropy for a partition $B$ of the system is $S=-\tr(\rho\log_2\rho)$. Therefore their entropies are
\begin{eqnarray}\label{entropARaR}
\nonumber S_{R}&=&-\sum_{n=0}^{N}\frac{\tanh^{2n}r}{C_N(r)^2\cosh^2r}\left[1+\frac{n}{\sinh^2r}\right]\\*
&&\nonumber\times\log_2\left[\frac{\tanh^{2n}r}{C_N(r)^2\cosh^2r }\left(1+\frac{n}{\sinh^2r}\right)\right]\\*
\nonumber S_{\bar R}&=&-\sum_{n=0}^{N-1}\frac{\tanh^{2n}r}{C_N(r)^2\cosh^2r}\left[1+\frac{n+1}{\cosh^2r}\right]\\*
&&\nonumber\times\log_2\left[\frac{\tanh^{2n}r}{C_N(r)^2\cosh^2r }\left(1+\frac{n+1}{\cosh^2r}\right)\right]\\*
\nonumber&&-\frac{\tanh^{2N}r}{C_N(r)^2\cosh^2r }\log_2\left(\frac{\tanh^{2N}r}{C_N(r)^2\cosh^2r}\right)\\*
\nonumber S_A&=&2\log_2\left[C_N(r)\right]-\frac{1}{C_N(r)^2}\sum_{i=0,1}D^i_N(r)\log_2\left[D^i_N(r)\right]\\*
\end{eqnarray}

We obtain a universal result which relates the entropies of the different bipartitions of the system, 
\begin{equation}
S_{AR}=S_{\bar R},\qquad S_{A\bar R}=S_R,\qquad S_{R\bar R}= S_A
\end{equation}  
These results can be summarised in the expression
\begin{equation}
S_{IJ}=S_K
\end{equation}
where $I,J$ and $K$ labels represent different subsystems. Whichever values $I\neq J\neq K$ will satisfy the identity. This is also true for standard scalar fields, Grassman scalar fields (spinless fermions) and Dirac fields. These relationships are completely universal, being independent of statistics and dimension, and reflect a fundamental aspect of Unruh decoherence in terms of the entropy of the partial systems, namely,  the way in which the entropy of the bipartitions behaves as acceleration increases  is not independent from the way the individual entropies do. 

The mutual information for all the possible bipartitions of the system will be
\begin{eqnarray}
\nonumber I_{AR}&=&S_A+S_R-S_{A R}=S_A+S_R -S_{\bar R}\\*
\nonumber I_{A\bar R}&=&S_A+S_R-S_{A\bar R}=S_A+ S_{\bar R}-S_R\\*
\nonumber I_{R\bar R}&=& S_R+S_{\bar R}-S_{R\bar R}=S_A+ S_{\bar R}+S_R
\end{eqnarray}  

The first notable difference from the standard bosonic field is that we do not obtain here the conservation law of the mutual information for the system Alice-Rob and Alice-AntiRob
\begin{equation}\label{noconservationbos}
 I_{AR} + I_{A\bar R}=2S_{A}
\end{equation}
And that for any finite $N$, $S_A$ goes to zero when $a\rightarrow\infty$. Only in the limit of $N\rightarrow\infty$ the conservation law is restored. 

\begin{figure}[h]
\includegraphics[width=.50\textwidth]{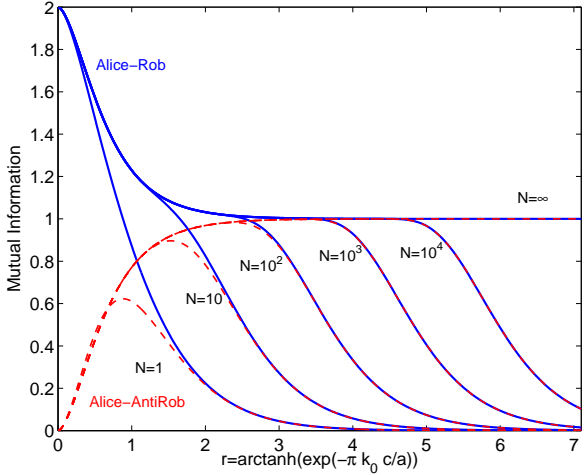}
\caption{(Colour online) Mutual information for the systems Alice-Rob (blue continuous lines) and Alice-AntiRob (red dashed lines) as the acceleration parameter varies. Several values of $N$ are plotted along with the $N=\infty$ case. A conservation law is satisfied until acceleration reaches a critical value which is displaced to the right logarithmically as $N$ increases.}
\label{MutuARAAR1}
\end{figure}

Figure \ref{MutuARAAR1} shows  the mutual information tradeoff of the systems $AR$ and $A\bar R$ from $N=1$ to $N=10^4$, along with the limit $N\rightarrow\infty$ in which the conservation law is fulfilled for all values of $a$. 
\begin{figure}[h]
\includegraphics[width=.50\textwidth]{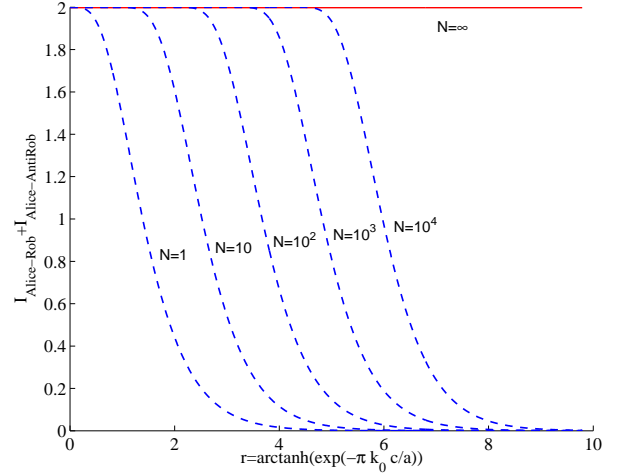}
\caption{(Colour online) Violation of mutual information conservation law for the systems $AR$ and $A\bar R$. The conservation law is fulfilled until $a$ reaches a critical value which is logarithmically displaced to the right as $N$ increases. The conservation law is completely restored when $N\rightarrow\infty$}
\label{conserva}
\end{figure}
\begin{figure}[h]
\includegraphics[width=.50\textwidth]{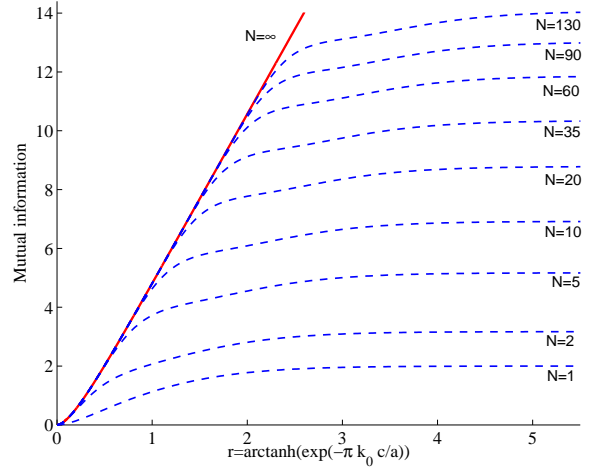}
\caption{(Colour online) Behaviour of the mutual information for the system Rob-AntiRob as acceleration varies for different values of $N$ showing the upper bound reached when $a\rightarrow\infty$. Mutual information diverges in the black hole limit only for $N\rightarrow\infty$.}
\label{mutuRARbos}
\end{figure}

As it can be seen in figure \ref{conserva} the largest deviation from the conservation law is obtained for $N=1$. As it is shown in the figure, for a given $N$ the conservation law is fulfilled until the acceleration reaches a critical value $a=a_{l}$, then correlations go rapidly to zero. This critical value increases logarithmically with $N$. 

We showed above that quantum correlations between $AR$ are quickly lost as $a$ increases for all $N$  and no entanglement is created between $A\bar R$. This means that for the high acceleration regime (where all quantum correlations vanish) classical correlations dominate mutual information. Therefore, what we learn from mutual information in this regime is the behaviour of purely classical correlations which are usually very difficult to be studied separately from the  quantum ones.

For the scalar field we have seen that, conversely to fermionic fields \cite{Edu3,Edu4},  limiting the dimension produces boundary effects which make classical correlations go to zero. Here, conservation of these correlations for all values of $a$ requires infinite dimension. In other words, finite dimensions schemes kill classical correlations as Rob accelerates.

One would expect something similar for fermions since their states are naturally of finite dimension. Hence, similar `border effects' in classical correlations should appear in the same fashion as for bosons. Despite this fact, mutual information for fermions does not vanish. The explanation for this difference between bosons and fermions comes from fermionic quantum correlations. As shown in \cite{Edu4}, there is a conservation law for fermionic quantum entanglement 
\begin{equation}\label{connetfer}
\mathcal{N}^{AR}_{\text{fermions}}+\mathcal{N}^{A \bar R}_{\text{fermions}}=\frac12
\end{equation}
Since classical correlations for the finite dimensional case go eventually to zero, quantum correlations rule mutual information behavior for fermions. Therefore, it can be concluded that  the conservation law for the mutual information for fermions must be strongly related with the conservation of the fermionic entanglement which has his origin in statistics. 

We can conclude then that the origin of the universal mutual information conservation law is different for fermions and bosons. On one hand, for bosons, it appears as a classical correlations conservation law. On the other hand, for fermions, this law reflects a quantum correlations conservation. This can also explain why mutual information behaves so similarly to negativity for fermions as it was obtained in \cite{Edu4}.  

To finish this work and complete the analysis of mutual information let us show in figure \ref{mutuRARbos}  how the behaviour of $I_{R\bar R}$ changes as $N$ is incresaed and how the divergent limit is obtained when $N\rightarrow\infty$. The results about mutual information here are coherent with the thorough analysis of the correlations for the $R\bar R$ bipartition performed above when we analysed negativity.

\section{Conclusions}\label{conclusions}

In this paper we answer the question of the actual impact of Fock space dimensionality on Unruh decoherence phenomena. To do so, we have studied the dimensional dependence of scalar field correlations when one observer is non-inertial. With this end in sight we have built a scalar field entangled state in which we have imposed a maximum occupation number $N$ for Rindler modes.

We have shown that the comportment of the entanglement for $AR$ and $A\bar R$ is only slightly influenced by $N$ so in other words, the qualitative behaviour (quick loss of entanglement for $AR$ as shown in figure \ref{bundle} and no entanglement creation for the system $A\bar R$) is the same for finite and infinite $N$. This again points to the argument in \cite{Edu3,Edu4} that it is statistics and not dimensionality that conditions the comportment of correlations in the presence of horizons.

However, we have shown that $AR$ entanglement is sensitive to variations of $N$. This opposes what we found for fermions, whose correlations are completely insensitive to Hilbert space dimension variations (for example going from Grassman scalars to Dirac fields).

In previous works we found a universal behaviour for the fermionic entanglement of the bipartition $AR$. Specifically the functional form for the negativity was exactly the same; independent of the maximally entangled state selected, the spin of the field, and the number of modes considered going beyond SMA. In all the cases Unruh decoherence degrades fermionic entanglement exactly the same way. Here we see that for bosons this universality principle does exist but it is not as strong due to the sensitivity of $AR$ to dimension changes. Extending this formalism to the electromagnetic field will reveal more information about this, thus it is expected to appear elsewhere.

We have also seen that lesser $N$ does not necessarily imply faster entanglement degradation. Instead, we have shown that for two different finite values of $N$, namely $N_1<N_2$ there is a region $a<a_c$ in which entanglement is more degraded for $N_2$ and another region $a>a_c$ in which entanglement is more degraded for $N_1$. In other words, for high accelerations, higher dimension means less entanglement degradation by Unruh effect. This result clashes again with the extended idea that lesser dimension would protect correlations better than higher dimension, one misconception that after all this research should be banished from the explanation of these phenomena.Ê 

We have also showed that, since $a_c$ shifts to the right as $N$ is increased, in the limit $N\rightarrow\infty$, $a_c\rightarrow\infty$, so that entanglement is more degraded for the infinite dimensional case than for any finite $N$ whatever the value of the acceleration.

It is remarkable that there is no entanglement creation in the CCA bipartitions even for finite dimension; no entanglement is created for the bipartition $A\bar R$ whatever the dimension limit $N$. This reflects again that the differences between fermions and bosons have nothing to do with the finite dimensionality, but with the different statistics. 

Concerning mutual information, we have shown that the conservation law found for scalar fields \cite{Edu4} for the systems $AR$ and $A\bar R$ is violated for finite values of $N$. We have obtained that for a finite $N$ the conservation law is fulfilled until acceleration reaches a critical value, in which correlations quickly drop. This critical value grows logarithmically with $N$, which means that the conservation law is satisfied for all $a$ when $N\rightarrow\infty$. Therefore, the violation of this conservation law can be associated with the boundary effects of imposing a dimensional limit. It is important to observe that in the bosonic case, mutual information is mainly accounting for classical correlations. This is because quantum correlations among these systems are quickly lost as $a$ increases.

Surprisingly, fermions, which have necessarily a limited dimension Hilbert space for each mode, lack these boundary effects. This big difference between fermions and bosons is related with the statistical conservation of quantum correlations in the black hole limit for fermions. Conversely to bosons, the high acceleration regime for fermions does not mean that mutual information is reflecting any classical correlations. Instead, mutual information is mainly reflecting Êfermionic quantum correlations. These correlations satisfy themselves a conservation law \eqref{connetfer} which is `inherited' by mutual information. This also explains the similitude between mutual information and negativity behaviour for fermions found in \cite{Edu4}.

The conclusion here is that the universal conservation law for mutual information is found for both fermions and bosons, however the nature of this conservation is completely different. With reference to bosons it is due to classical correlations conservation, whereas for fermions it is due to quantum correlations conservation. This illustrates yet again that statistics is such a paramount feature in order to explain how correlations behave in the presence of horizons.

The dimension of the Fock space has the largest impact in the behaviour of correlations between regions $I$ and $IV$ separated by the horizon. Comparing the limited dimension scalar fields with their fermionic analogs we have found that the behaviour of correlations $R\bar R$ is somewhat similar for bosons and fermions and mainly ruled by the dimensionality of the Hilbert space. For both fermionic and bosonic fields, entanglement is always created between $R\bar R$ reaching a maximum value when $a\rightarrow\infty$ for any finite dimension.

The scalar cases of finite $N$ present, however, important differences compared with their fermionic Fock space dimensional analogs (namely $N=1$ is analogous to the Grassman scalar case and $N=2,4$ to the Dirac field case). In the black hole limit, scalar states entanglement created between modes in both sides of the horizon is greater than in the fermionic case.

The scalar cases of finite $N$ however present important differences compared with their fermionic Fock space dimensional analogs (namely $N=1$ is analogous to the Grassman scalar case and $N=2,4$ to the Dirac field case). In the black hole limit the scalar states entanglement created between modes in both sides of the horizon is greater than in the fermionic case.

This implies that, for correlations between modes of Rob and AntiRob, the effect of statistics in the black hole limit is the opposite to what happened for the CCA bipartitions. Here bosonic correlations reach higher values than their corresponding fermionic analogs.

\section{Acknowledgments}

This work was partially supported by the Spanish MICINN Project FIS2008-05705/FIS and by the CAM research consortium QUITEMAD S2009/ESP-1594. E. M-M was partially supported by a CSIC JAE-PREDOC2007 Grant.

\bibliographystyle{apsrev}

\end{document}